\documentclass[fleqn,10pt]{wlscirep}
\usepackage[utf8]{inputenc}
\usepackage[T1]{fontenc}
\usepackage{tabto}
\usepackage{multirow}
\usepackage{tabularx}
\usepackage{cite}
\usepackage{caption}
\usepackage{subcaption}
\usepackage{amsmath,amssymb,amsfonts}
\usepackage[font=small,skip=0pt]{caption}
\usepackage{algorithmic}
\usepackage{textcomp}
\usepackage{subcaption}
\usepackage{xcolor}
\usepackage{balance}
\usepackage{tabto}
\usepackage{float}
\usepackage{subcaption}
\usepackage{hyperref}
\usepackage{adjustbox}
\usepackage{dblfloatfix}
\usepackage{graphicx}
\usepackage{hyperref}
\usepackage{tabularx}
\usepackage{makecell}
\usepackage{enumitem}
\title{Towards More Transparent and Accurate Cancer Diagnosis with an Unsupervised CAE Approach}

\author[1,2,*]{Zahra Tabatabaei}
\author[2,3]{Adrián Colomer}
\author[1]{Javier Oliver Moll}
\author[2]{Valery Naranjo}
\affil[1]{Dept. of Artificial Intelligence, Tyris Tech S.L.,Valencia,46021, Spain.}
\affil[2]{Instituto Universitario de Investigación \\ en Tecnología Centrada en el Ser Humano,\\ HUMAN-tech, Universitat Politècnica de València, Spain.}
\affil[3]{ValgrAI – Valencian Graduate School and Research Network for Artificial Intelligence}
\affil[*]{elec.tabatabaei@gmail.com, zahra.tabatabaei@tyris-software.com}


\begin{abstract}
\textit {Background:}
Digital pathology has significantly impacted the cancer diagnosis field, with Content-Based Medical Image Retrieval (CBMIR) emerging as a powerful tool for analyzing histopathological Whole Slide Images (WSIs). CBMIR allows users to search a database for similar content to a query, providing pathologists with access to collections of cases with comparable features. This can improve the reliability of diagnostic references and help in making more accurate and timely diagnoses.

\textit {Objective:}
In 2020, the Global Cancer Observatory (GCO) reported that breast cancer is the most prevalent cancer type in both men and women, accounting for 11.7\% of all cases, while prostate cancer is the second most common cancer type in men, comprising 14.1\% of cases. The aim of the proposed Unsupervised CBMIR (UCBMIR) is to replicate the traditional cancer diagnosis workflow and provide a dependable method for supporting pathologists when making diagnostic conclusions based on WSIs. By reducing the workload of pathologists, this approach could potentially enhance the accuracy and efficiency of cancer diagnosis.

\textit {Method and results:}
The study presents an innovative approach to address the problem of the lack of labeled histopathological images in CBMIR. A customized unsupervised Convolutional Auto Encoder (CAE) was developed to extract $200$ features per image, which were then used by the search engine component. The proposed UCBMIR was evaluated using two widely used numerical techniques in CBMIR and visual evaluation, and compared with a classifier to determine if retrieved images belong to the same cancer type as the query. The validation process was conducted using three distinct data sets, with an external evaluation to demonstrate its effectiveness. The UCBMIR outperformed previous studies, achieving a top 5 recall of $99\%$ and $80\%$ on BreaKHis and SICAPv2, respectively, using the first evaluation technique. Using the second evaluation technique, UCBMIR achieved precision rates of $91\%$ and $70\%$ for BreaKHis and SICAPv2, respectively. Moreover, the UCBMIR was able to identify various patterns in patches and achieved an accuracy of $81\%$ in the top 5 when tested on an external image from Arvaniti, having been trained using SICAPv2 with the second evaluation technique.

\end{abstract}
\begin{document}

\flushbottom
\maketitle
%
%
\thispagestyle{empty}

\section*{Introduction}\label{sec:introduction}
Cancer is a leading cause of death worldwide, with nearly 10 million deaths reported in 2020, as per the World Health Organization (WHO) \cite{pineros2021scaling}. In 2020, breast and prostate cancer affected 2.26 million and 1.41 million cases, respectively. Accurate cancer diagnosis is critical for effective treatment because each cancer type requires a specific treatment regimen. However, diagnostic errors are prevalent, affecting approximately 5.08\% of cases, which translates to around 12 million adults in the United States \cite{singh2014frequency}. This significant percentage of human error in a large number of cancer cases poses significant drawbacks for society and the quality of human lives.

Moreover, there is a significant disparity in treatment availability between countries with varying income levels. In high-income countries, comprehensive treatment is available in over 90\% of cases, but this figure drops to less than 15\% in low-income countries \cite{world2019global}. In this context, there is an urgent need to develop reliable and accurate diagnostic tools that can assist medical professionals in making accurate and timely diagnoses, regardless of their location or income level.

Computer-Aided Diagnosis (CAD) models play a vital role in reducing the incidence of human errors and providing an inclusive worldwide platform for individuals with varying incomes. CAD offers multiple approaches under the umbrella of "digital pathology" to enhance conventional cancer diagnosis. Digital pathology has garnered significant attention due to its ability to provide a definitive diagnosis at the pathology level, taking into account factors such as size, complexity, and color \cite{kalra2020yottixel}. The challenges and opportunities presented in digital pathology are explained in \cite{tizhoosh2018artificial}. Despite the challenges, digital pathology can serve as a bridge toward the discovery of histopathological imaging and enable more accurate prognostic predictions for disease aggressiveness and patient outcomes.

The following subsections cover a brief literature review of digital pathology on WSIs. 
\subsection{Segmentation}\label{Segmentation}
Automatic detection of irrelevant regions of tissue may bring a more reliable prediction. In \cite{fuster2022invasive}, they proposed a multi-scale model to detect invasive cancerous area patterns in WSIs of bladder cancer. Similarly, \cite{kanwal2022quantifying} focuses on detecting blood and damaged tissue as problematic artifacts in bladder tumors.  
In \cite{kiran2022denseres}, the authors apply the DenseRes-Unet model to multi-organ histopathological images to segment overlapped/clustered nuclei. A binary threshold is set to detect the contour of the extracted nuclei in the images, as the morphological characteristics of the cells are critical to grading the cancers. Moreover, the two-stage nuclei segmentation strategy proposed in \cite{hu2022breast} based on watershed segmentation is used to distinguish between carcinoma and non-carcinoma recognition in the Bio-imaging 2015 data set. 
Additionally, \cite{xu2015stacked} introduced a novel approach to detect nuclei in breast cancer histopathological images using a stacked sparse Auto Encoder (AE). 

Segmentation techniques have been studied extensively to quantify cell nucleus form and dispersion, which may improve accuracy in classification and grading \cite{li2018path}. However, these methods do not offer direct benefits to pathologists. Though Deep Learning (DL) has shown promise in improving segmentation, it relies heavily on large annotated data sets \cite{litjens2017survey}, limiting its impact \cite{li2022high}. Innovative approaches are needed to develop new techniques that can benefit pathologists and improve disease diagnosis.

\subsection{Classification}\label{Classification}
Classification of input images is a critical task in medical image analysis, where an optimal classifier is expected to provide accurate labels for each input \cite{tabatabaei2023selfsupervised}. This can significantly aid pathologists in their daily analysis of tissue grading. For instance, \cite{going} validated an end-to-end pixel-level prediction of Gleason grades and scored the entire biopsy. Similarly, \cite{liu2022breast} proposed an AE using Siamese network aimed at learning image features by minimizing the distance between input and output. Another approach was proposed by \cite{ahmad2022transfer}, who aimed to decrease the rate of diagnostic errors by performing patch-based transfer learning. However, patch-level data sets extracted from Whole Slide Images (WSIs) often contain mislabeled patches, which may lead the classifiers to miss important information. To address this problem, \cite{9177091} proposed DenseNet121-AnoGAN, which employs unsupervised anomaly detection with generative adversarial networks (AnoGAN) to prevent missing mislabeled patches. This approach has been successfully applied to classify the BreaKHis data set into benign and malignant.

The author in \cite{alom2019breast} fed a Inception Recurrent Residual Convolutional Neural Network (IRRCNN) model with two breast cancer data sets to have binary and multi-class classifier.
The authors in \cite{li2020classification} conducted their experiments on DenseNet with SENet IDSNet and BreaKHis data set. They fine-tuned DenseNet-121 to propose an accurate classifier. 
A deep Feature Extractor (FE) from a pre-trained network and a classifier are used in \cite{abbasniya2022classification} to classify BreaKHis. 16 pre-trained networks and 7 classifiers were tested in this paper.

In brief, classifier architectures have been proposed for use in diagnostic pathology to aid pathologists in making more accurate cancer diagnoses. Many studies have reported high classification accuracy, which has been validated in engineering laboratories. However, despite their potential importance, these measurements have not yet led to a significant change in diagnostic imaging. 

While classification and segmentation have proven to be valuable tools, they have not drastically transformed the diagnostic process. This may be attributed to their inability to reduce ambiguity and boost the confidence of pathologists in their diagnoses. In essence, these methods do not provide any additional information to aid pathologists in their report writing during the diagnostic process.

In regards to developing methods for specific applications, it is possible to achieve higher results for the intended objective. However, creating and implementing unique methods for each potential task of interest is impractical. As an alternative approach, CBMIR has established a reliable framework for quality control. While it may have poorer accuracy than an application-specific instrument, having a multipurpose general-purpose tool like CBMIR can still be useful.

\subsection{Content-Based Medical Image Retrieval (CBMIR)} \label{CBMIR}

Automated medical imaging has been growing dramatically to improve clinical treatment and intervention in medical diagnosis. This yields an exigent demand for developing highly effective CAD systems. CBMIR is an active area of research with significant applications in routine clinical diagnostic aid, medical education, and research. In CBMIR, the end user targets retrieving the most relevant images. So, pathologists will trust this outcome easier because not only will they have a second opinion on their tissue (label), but they can also look for the same patterns in the previous tissues. In the classification task, they can get a label for their new tissue without knowing the reason. But in CBMIR, they can see the similarity between their tissues and the retrieved ones. Moreover, the explainable nature of CBMIR allows clinicians to understand how the system arrived at a particular diagnosis or recommendation, promoting transparency and trust in AI-assisted medical decision-making. Most notably, CBMIR is pathologist-centric; in contrast to classification, it is essentially an attempt to make decisions on behalf of the pathologists.

In DL, similar patterns mean similar features and representations. Humans can properly describe and interpret image contents, while digital machines can provide fewer semantic words for the same image. Machines provide a numerical description of the images with a wide gap compared to the human interpretation of the same image. This gap is named "\textit{semantic gap}," and this broadly limits the performance of retrieval tasks \cite{taheri2022effective}. The semantic gap is the main reason CBMIR has not made it into the daily laboratories workflow, yet. Indeed, this is arguably the paramount challenge in adopting CBMIR into the laboratories' workflow. Pathologists face numerous challenges in the current diagnostic paradigm, with time being a common factor. However, the impact of these challenges extends beyond just medical professionals and patients; it can also affect society as a whole. This can lead to emotional distress and other adverse effects on the well-being of patients and their families. Digital pathology, through the use of CBMIR, can mitigate the impact of these changes and enhance the accuracy of diagnoses.

CBMIR in virtual telepathology offers a reliable framework for achieving quality control through computational consensus-building, ensuring that diagnoses are accurate and consistent across different pathologists and healthcare institutions. By utilizing a vast database of reference images and advanced algorithms, CBMIR enhances the accuracy of diagnoses, potentially decreasing the need for additional studies and speeding up the diagnostic process. This can lead to better patient outcomes and a more efficient healthcare system. In recent years, CBMIR has gone through a renaissance with the promise of revolution.
In a previous study \cite{9502205}, a CNN-based AE was applied to the BreaKHis data set with the aim of minimizing misinformation and evaluating the performance of CBMIR in a binary scenario. However, the reconstructed images produced by this method were found to be blurry, indicating that the extracted features by the AE were not robust enough to reconstruct the original image. In addition, the scope of this study was limited to detecting breast cancer using a two-class data set, without considering other diseases. These limitations highlight the need for further research to improve the quality of feature extraction in CBMIR systems. In \cite{mazaheri4216426ranking}, the CBMIR performance was improved in a supervised manner using a Hybrid feature-based ICNN model. The model was trained by adding three Fully Connected (FC) layers to accommodate the classification of cancer subtypes from TCGA. The researchers in \cite{rasoolijaberi2022multi} aimed to replicate the process of detecting morphological features used by pathologists in cancer diagnosis by incorporating different magnification levels into their CBMIR system. Specifically, they trained their system using a subset of TCGA data set in three magnification levels: 20$\times$, 10$\times$, and 5$\times$. To address the differences in features that might exist at these different magnification levels, the last DenseNet-121 block \cite{iandola2014densenet} was re-trained using 10$\times$ and 5$\times$ magnification patches. This supervised approach improved the adaptability of the FE and resulted in better overall performance of the CBMIR system. KimiaNet reported two types of image search: horizontal search and vertical search. In the horizontal search, the query is applied to the entire data set to find similar whole slide images (WSI) with a self-supervised model \cite{kalra2020pan}, while in the study by Fashi et al. \cite{fashi2022self}, the vertical search approach is designed to identify similar types of malignancies in a specific organ. This is achieved by utilizing pretrained models with openly provided weights from the Keras library. The problem with supervised CBMIR is that it requires a large amount of labeled data, which can be time-consuming and costly to obtain. On the other hand, the problem with self-supervised CBMIR is that it may not perform as well as supervised methods and may require more complex models. Indeed, many researches 
\cite{zheng2003design,shi2017supervised,pantanowitz2021digital,caicedo2011content,qi2014content,sridhar2015content,kwak2016automated,zhu2018multiple,babaie2017classification,zhang2014towards} have been dedicated to CBMIR, but the overall performance of the existing systems is not high enough due to the growing medical images and digital pathology.

The main contributions of this paper in proposing an Unsupervised CBMIR (UCBMIR) are:

\begin{itemize}
\item Proposing a new unsupervised approach for prostate gradation problem using CBMIR that achieves performance comparable to fully supervised methods on the largest pixel-wise annotated prostate data set.
\item Extensively validating the proposed UCBMIR approach on three databases, including BreaKHis for binary and SICAPv2 and Arvaniti for multi-class, which is more challenging.
\item Conducting an external evaluation to demonstrate the performance and generalization of UCBMIR, by training the model on SICAPv2 and testing it on the Arvaniti data set.
\item A comprehensive evaluation of UCBMIR is presented, encompassing various numerical and visual performance metrics.
\end{itemize}
In addition, the paper addresses two major problems in traditional cancer diagnosis: inexperienced pathologists requiring more ancillary studies for diagnosis and the time-consuming process of differentiating between cancer grades. The UCBMIR model proposed in the paper provides a vast database of images that pathologists can use as a reference for diagnosis, allowing them to make more accurate diagnoses even if they are inexperienced. Additionally, the proposed tool enables pathologists to access annotated image databases instantly, leading to a faster diagnosis and skipping time-consuming reading and searching processes in "\textit{ExpertPath}" and "\textit{PathologyOutlines}" or an Atlas book.


In order to reach the primary objective of this study we introduce an unsupervised image search tool for pathologists to facilitate the efficient retrieval of similar images from previous cases. The initial stage involves training a customized CAE that includes a skip layer between the encoder and the decoder, as well as an attention block in the bottleneck. This CAE is trained to reconstruct images and learn effective data representations while simultaneously ignoring the noise. The encoder with the bottleneck of the trained CAE serves as our FE in the search stage. We represent the complete training set of the data set as in previous cases and carry out patch-by-patch retrieval to obtain diagnosis-relevant patches for each query in the test set. Our ranking algorithm, which utilizes Euclidean distance, identifies the retrieved patches, which are then presented to the pathologists as the output of the proposed UCBMIR. Our study showcases the practicality of our approach in enhancing the efficiency and accuracy of image retrieval for pathologists and engineers. As a result, our method can accelerate cancer diagnosis for pathologists, and the deep layers in the custom-built CAE can learn image features in an unsupervised manner, circumventing the issue of insufficient training images.

\section{Material}
The study evaluates the performance of the UCBMIR on two of the largest labeled histopathological images in breast and prostate cancer, namely BreaKHis and SICAPv2, respectively. The Arvaniti data set is utilized as the third and an external data set in order to validate the model performance. These two cancers, prostate and breast cancer, are selected as they are prevalent in society.

\textbf{BreaKHis:}
breast tissue biopsy slides were stained with Hematoxylin and Eosin (H \& E) and labeled by pathologists at the P\&D medical laboratory in Brazil \cite{spanhol2015dataset}. This data set is composed of $7909$ microscopic images of breast tumor tissues collected from patients using magnifying factors of $40\times$, $100\times$, $200\times$, and $400\times$ in the size of $224\times224\times3$. This binary data set contains $588$ benign and $1232$ malignant images in $400\times$. 

\textbf{SICAPv2:}
prostate samples were sliced, stained in H \&E, and digitized at $40\times$ magnification. Images were divided into $512\times512\times$3 and down-sampled to $10\times$, which is commonly used for evaluating images. This multi-class data set contains $155$ WSIs in total: $4417$ non-cancerous patches, of which $1635$ are labeled as Grade 3 (G3), $3622$ as Grade 4 (G4), and $665$ as Grade 5 (G5), Table \ref{tab1:SICAPv2 data set description.}. Images labeled by a group of expert urogenital pathologists at Hospital Clínico of Valencia . SICAPv2 is the largest publicly available data set that includes pixel-level annotations of Gleason grading, providing detailed information on the presence of cribriform patterns\cite{going}.

\begin{table}[ht]
\caption{SICAPv2 data set description.}
\begin{center}
\begin{tabular}{|c|c|c|c|c|}
\hline
\textbf{Grades}&\textbf{NC} & \textbf{G3} & \textbf{G4} & \textbf{G5} \\
\cline{1-4} 
\hline\hline
WSIs & 37 & 60 & 69 & 16\\
\hline
Patches & 4417 & 1636 & 3622 & 655\\
\hline
\end{tabular}
\label{tab1:SICAPv2 data set description.}
\end{center}
\end{table}

In order to validate the generalization capability of the UCBMIR to find similar images, Arvaniti, an external data set containing pixel-level annotations of Gleason grades, is used. 

\textbf{Arvaniti:}
the data set was shared by Arvaniti et al. \cite{arvaniti2018automated}, which contains $625$ patches of prostate histology images at $40\times$ magnification. Regarding a fair comparison with SICAPv2, in \cite{going}, some configurations were applied to re-sample images to $512\times512$ at $10\times$ magnification. To normalize the color distribution of Arvaniti, the author in \cite{going} applied a histogram match to the re-sampled images and set the images in SICAP and Panda\cite{bulten2022artificial} as the reference images. These re-sampled images are used as the third data set and the external evaluations in this paper. Arvaniti is employed for performance evaluation alongside normalization by both Panda and SICAP, in addition to the external evaluation. This is discussed in detail in the following sections of this journal paper.
\section{Methodology}
A CBMIR contains four subsections: 1. training, 2. indexing and saving, 3. searching, and 4. evaluating. The search tool in CBMIR uses the contents within each pixel of the images instead of using annotations or metadata. Consequently, similar images are retrieved from a large data set that matches the contents of the queried image. Also, it is often impractical to manually annotate images in a large data set, thus an unsupervised FE is developed in this study to address this issue.

The four phases of the proposed UCBMIR are described in-depth in the following subsections with Figure \ref{fig:y equals x} and Figure \ref{fig:three sin x}, in accordance with SICAPv2, as it is a complex multi-class data set.
\subsection{Training}
The only training part of the proposed UCBMIR is training the proposed CAE. CAE aims to reconstruct the output as equal to the input. CAE could learn effective features with unlabeled data in an unsupervised manner. Using a CAE approach offers several benefits, such as its ability to capture spatial information through convolutional layers, which are well-suited for processing image data. Moreover, CAE employs multiple layers to extract advanced image representations, resulting in higher-level feature recognition. These multi-layered models have fewer free parameters, making them simpler and faster to train, reducing the cost and resources required for training. Figure \ref{fig:y equals x} exhibits an illustration of the proposed CAE architecture. It contains three main parts:

\begin{figure}[b!]
\centerline{\includegraphics[width=0.48\textwidth]{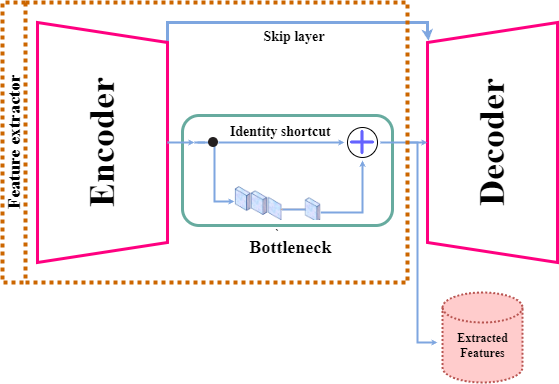}}
\caption{ proposed CAE architecture with kernel size of $3$ throughout the model, the stride of $2$ in the encoder and decoder, and $1$ in the bottleneck layer.}
\label{fig:y equals x}
\end{figure}

\begin{figure}[b!]
\centerline{\includegraphics[width=0.95\textwidth]{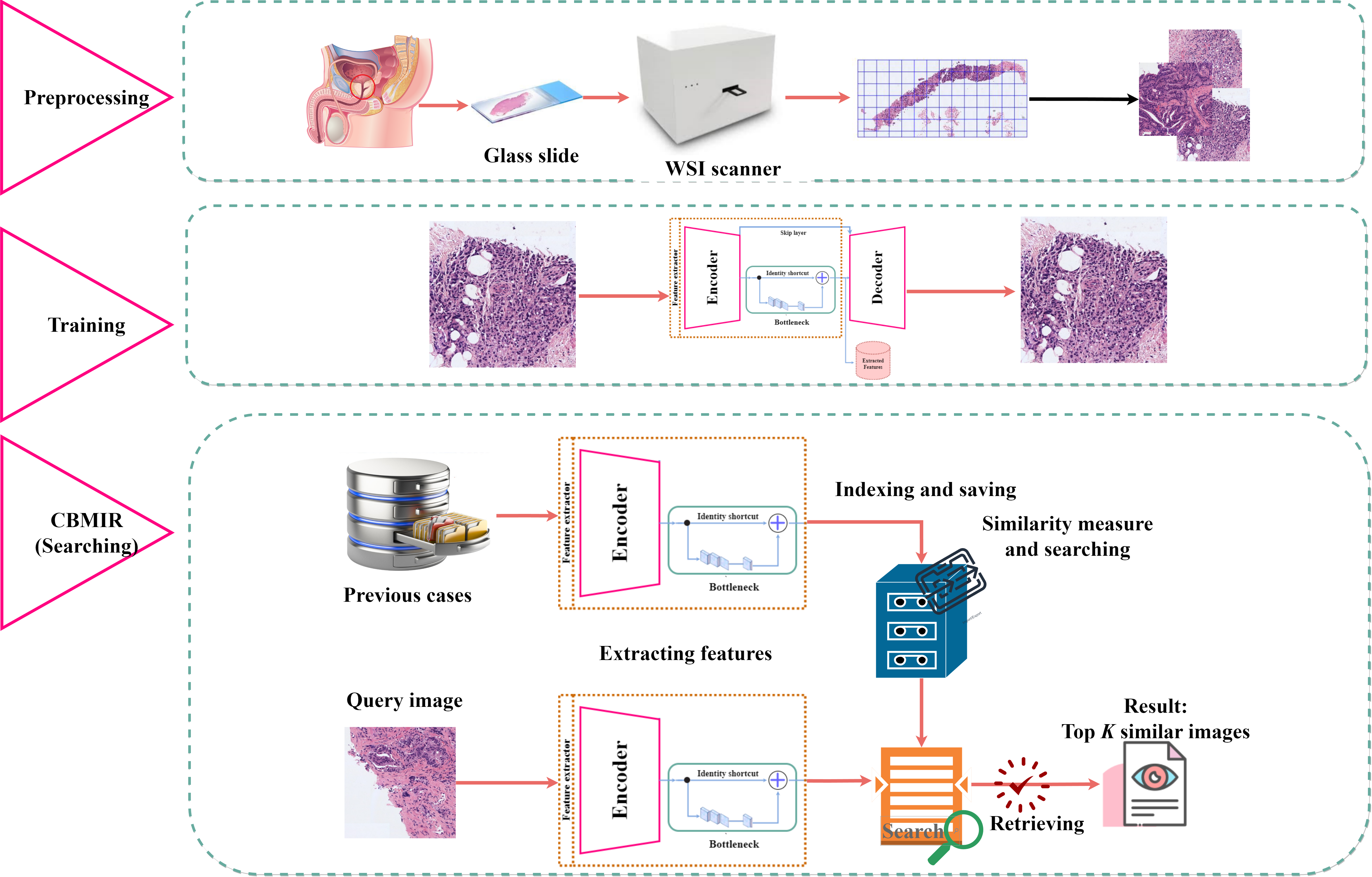}}
\caption{an overview of the UCBMIR for retrieving similar cases to a given query. The preprocessing stage involves extracting tissue from the patient's body and dividing the whole slide images (WSIs) into patches of a specific size\cite{going}. In the training stage, these patches are used to train the proposed unsupervised CAE to extract feature representations. The trained encoder and bottleneck layers are then used to extract feature embeddings FEs that are used in the CBMIR section. In the CBMIR stage, the search engine computes the embedding features of the training set and stores them in a dictionary. When a query image is selected from the test set, the FE computes the embedding of that query and compares it with those in the dictionary. The model then returns the \textit{K} most similar patches based on the pathologists' needs. }
\label{fig:three sin x}
\end{figure}

\begin{itemize}
    \item \textbf{Encoder}:
        it captures the structural attributes of the input images across a feature vector per image with $200$ elements. The sizes of the convolutional filters in the encoder are $[16, 32, 64, 128, 256]$. Histopathological images are highly detailed; using operations such as pooling layers will cause them to lose lots of information. Because of that, in the proposed CAE, the size of the images decreases by passing through convolutional layers without any pooling layer. 
    \item \textbf{Bottleneck}:
        it contains $200$ extracted features per image. As can be seen in Figure \ref{fig:three sin x} a residual block in the bottleneck contains four filters in the size of $[64, 32, 1, 256]$. 
    \item \textbf{Decoder}:
        it reconstructs the input from its $200$ intermediate feature vectors. Consequently, to the encoder, the filters in the decoder part are in the size of $[128, 64, 32, 16, 3]$.
\end{itemize}

The main objective of the proposed CAE is to find the most discriminative feature vectors to describe the images unsupervised. Briefly, it compresses input image patches (of dimensions width $\times$ height $\times$ channels) into a fixed-length vector. The performance of the FE is directly related to the depth of the learning model, as deeper and more complex models might result in overfitting. To address this issue, the proposed CAE employs a residual block in the bottleneck to increase the network depth and improve end-to-end mapping.

To get better performance gain, inspired by highway networks\cite{srivastava2015training} and deep residual networks\cite{he2016deep}, we add a skip connection between two corresponding convolutional and deconvolutional layers. When the network goes deeper, image details can be lost, making deconvolution weaker at recovering the input. Skip connections benefit by back-propagating the gradient to the bottom layers, making training a deeper network much more accessible. In other words, skip connections pass gradients backward, which helps find a better local minimum. 

Mean Square Error (MSE) is the loss function that updates the network weights during the training phase. The minimum amount of MSE means more similarity between the input and the output. The greater the similarity between the input and the reconstructed images, the more meaningful features are in the bottleneck. 
In practice, we find that using Adam with a learning rate $ 5 \times 10^{(-5)} $ can train the model in $10$ epochs. Then, the decoder part of the trained CAE is discarded, and the remaining sections, including the encoder and the bottleneck, play a role as a FE. 
\subsection{Indexing and saving}
The indexing and saving stage is a crucial step in CBMIR as it enables efficient storage and organization of extracted features. This, in turn, enables fast and accurate retrieval of relevant medical images during the search stage, thus improving the diagnosis and treatment of medical conditions. In our study, we used $n$ images from both the validation and train sets of each data set as input to the FE, resulting in $n$ feature vectors that represent each image in a $200-$dimensional latent space. These feature vectors were then stored in a dictionary, $D_i = [F_1, F_2, ..., F_n]$, where each $F_i$ contains the feature vectors for a single image.

During the retrieval stage, we utilized this dictionary as a reference for comparison with the query image. For the SICAPv2 data set, there were $2122$ query images. By organizing and storing the extracted features in this manner, we aimed to improve the efficiency and accuracy of our CBMIR system. This approach enabled us to retrieve medical images that were relevant to a given query, thereby aiding in faster and more accurate diagnosis and treatment of medical conditions. Figure \ref{fig:three sin x} illustrates the process of organizing and storing the extracted features for efficient retrieval of medical images during the search stage. 

\subsection{Searching}
The searching process in CBMIR involves three key steps: similarity calculation, ranking and retrieval, and visualization and presentation. During similarity calculation, the search engine uses similarity measures such as Euclidean distance, cosine, Manhattan, and Haversine to calculate the similarity between the query image and other images in the database. The images in the database are then ranked based on their similarity to the query image, and the top-ranked images are retrieved and presented to the user for further analysis.

In this paper, we experimented with both Cosine and Euclidean distances, and based on our results, we concluded that the Euclidean distance was the more suitable choice. We use Euclidean Distance to measure the similarity of two feature vectors. Specifically, we calculate the distance by each query feature $F_\textit{Q}$ with all the feature vectors in $D_i$, and the smaller Euclidean value corresponds to more similar images. Our experimental findings suggest that Euclidean Distance is an effective metric for measuring similarity in CBMIR systems. By accurately measuring the similarity between images, the search engine can more effectively retrieve relevant medical images, leading to improved diagnosis and treatment outcomes.

\subsection{Evaluation}
It is worth considering what "accuracy" means in the context of a CBMIR. The accuracy of CBMIR depends on what we are looking for and what is displayed by the search engine. The use case determines whether the search is looking for images with the same stain, comparable stain intensity, same histologic feature, or similar grade; hence, this objective is ambiguous. To address this lack of awareness of the intent of the search engine, top \textit{K} score at retrieving images of the same histologic features and Gleason grades engaged in the prior research to determine the performance of their experiments. 
To the best of the author's knowledge, there are two most-used strategies for calculating the top \textit{K} score described in the recent articles:
\begin{enumerate}
    \item If there is only one correct retrieved image, this has been shown as a correct answer \cite{tabatabaei2022residual}. In this paper, we set $\textit{K} = 3, 5, 7$, which evaluates the performance of our model to correctly present at least one correct result in the top \textit{K} retrieved images. In this paper, we name this method as "EV 1" regarding report the results in the following tables.
    \begin{equation}
        ACC@\textit{K} = \frac{1}{N}\sum_{i}^N \varepsilon(\alpha_i,TOP(ans[:K]))
    \end{equation}
    In this equation, $N$ denotes the number of query patches, and $\alpha_i$ represents the label of the $i$-th query patch. The function $TOP(ans_i[:K])$ retrieves the top $k$ most similar results for the query, and outputs $1$ if any of these results match with the query, and $0$ otherwise. In other words, if $TOP(ans_i[:K])$ belongs to the set of labels of the $i$-th query, denoted by $\alpha_i$, the function $\varepsilon()$ returns $1$.
    
    \item Precision \eqref{eq:eq1} and recall \eqref{eq:eq2} are the two selected indicators to evaluate the results. In this study, this is termed "EV 2".
    \begin{equation}
       Precision = \frac{R_v}{n} \label{eq:eq1}
    \end{equation}
    \begin{equation}
        Recall = \frac{R_v}{M}\label{eq:eq2}
    \end{equation}
\end{enumerate}

Herein, $R_v$ denotes the set of retrieved images that are considered relevant, while $n$ signifies the total number of captured images. Moreover, the number of relevant images present in the data set is explicitly annotated as $M$. 
The proposed UCBMIR is evaluated based on both top-ranking image retrieval strategies to mimic the standard search process.

\section{Discussion and results}

Matching pairs to the image is the core of any search engine, in which an image is compared to a database to determine similarities. Numerous studies have been conducted on CBMIR in a binary manner, as it is more challenging with multi-class data sets. 

Breast cancer is a prevalent malignancy affecting women globally. In the domain of CAD, the BreaKHis data set is a popular choice for evaluating the performance of algorithms in CBMIR. In this study, we employed the BreaKHis to assess the efficacy of our UCBMIR approach. Our method demonstrated superior performance in matching image pairs, as evidenced by the results presented in Table \ref{tab:Breakhis}. Specifically, our approach outperformed two previously reported methods, namely \cite{9502205} and \cite{gu2019multi}, with precision scores of $92\%$ and $91\%$, respectively, for both evaluation criteria (EV1 and EV2). These results suggest that our approach is highly effective in accurately identifying patterns in breast cancer images.

\setlength{\arrayrulewidth}{.1em}
\bigskip
\begin{table}[htp!]
\begin{center}
\vspace{-4mm}
\caption{comparative results on BreaKHis 400$\times$ at $\textit{K} = 5 $. We measure the precision and recall with both EV 1 and EV 2.}
\label{tab:Breakhis}
\begin{tabular}{|c|c|c|c|c|}
\hline
\textbf{\thead{Type of \\evaluation}}&\textbf{Method}&\textbf{Precision}&\textbf{Recall}\\
\cline{1-4} 
\hline\hline
\textit{EV 1}& UCBMIR & \textbf{0.92} & \textbf{0.99} \\
\hline
                & UCBMIR & \textbf{0.91} & \textbf{0.50} \\
\cline{2-4} 
                \textit{EV 2} & Minarno \cite{9502205} & 0.70 & 0.31  \\
\cline{2-4} 
                 & Gu \cite{gu2019multi} & 0.63 & -  \\
\hline

\end{tabular}
\end{center}
\end{table}

\begin{table*}[b!]
\begin{center}
\vspace{-4mm}
\caption{model quality results on SICAPv2 with top 5 retrieved images. The reported results are obtained by EV 1. The metrics are precision, recall, and accuracy.}
\label{tab:Prostate}
\begin{tabular}{lr<{\ }cccc}
\hline
 \textbf{Method} &\textbf{Data set} &\textbf{Precision} &\textbf{Recall} &\textbf{Accuracy}\\
\cline{1-5} 
\hline\hline
                UCBMIR & SICAPv2 & 0.79  & \textbf{0.80} & \textbf{0.79} \\

\cline{1-5} 
                UCBMIR & \thead{Patches normalized \\ Arvaniti (SICAP)} & 0.71 & 0.75 & 0.80\\
\cline{1-5} 
                 UCBMIR &\thead{Patches normalized \\ Arvaniti  (Panda)} & \textbf{0.80} & 0.68 & 0.78\\

\cline{1-5} 
                 \thead {Hegde \cite{hegde2019similar}\\  \textit{(SMILY)} }& TCGA & - & - & 0.73 \\
\cline{1-5}
                  \thead{Hegde \cite{hegde2019similar}\\  \textit{(SIFT)}}
                  & TCGA & - & - & 0.62 \\
\cline{1-5}
                 \thead{VGG16 \\ \textit{(ImageNet)})} & SICAPv2 & \textbf{0.80} & 0.80 & 0.80\\

\hline
\end{tabular}
\end{center}
\end{table*} 
In order to evaluate the effectiveness of our CBMIR method on a multi-class data set, we utilized SICAPv2 and Arvaniti data sets, both containing four classes. Given the global prevalence of prostate cancer, we selected this type of cancer for our experiments and used SICAPv2 as the largest pixel-wise annotated data set. The Arvaniti data set was re-sampled by referencing Panda and SICAP, as stated in \cite{going}, and was used in two experiments of this paper to demonstrate the robustness of our methodology.

Table \ref{tab:Prostate} presents the results obtained using our approach with $K=5$ and EV1 as the evaluation criteria. To demonstrate the efficacy of our methodology, we conducted two experiments using the Arvaniti data set. In the first experiment, to ensure a fair comparison, we trained the model using Arvaniti normalized based on both SICAP and Panda. The results of these two trained models, obtained by conducting the entire training and searching steps, are reported in Table \ref{tab:Prostate}. These findings demonstrate the superior performance of our approach in accurately identifying and classifying prostate cancer images in multi-class data sets, thereby potentially contributing to the development of improved diagnostic tools and clinical decision-making processes.

\begin{figure*}[htp!]
\centerline{\includegraphics[width=0.65\textwidth]{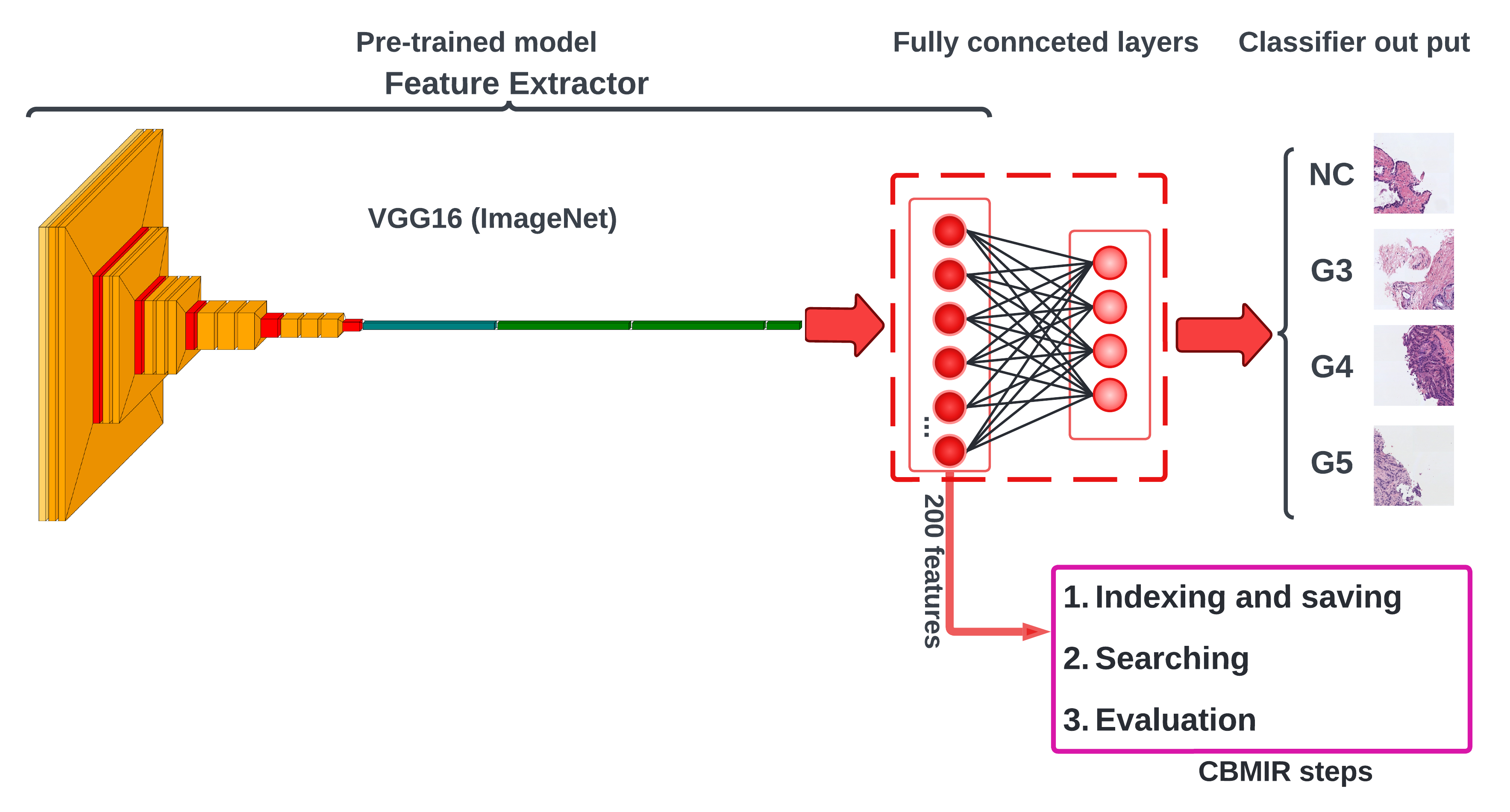}}
\caption{the structure of training VGG as a multi-class classification on SICAPv2 and delivering $200$ features per image to the following CBMIR steps. The Conv2D layers are shown in "orange", MaxPooling2D in "red", Dense layers in "green", and Flatten in "teal". }
\label{fig:VGG.}
\end{figure*}

In a study by Hegde \cite{hegde2019similar}, Scale-Invariant Feature Transform (SIFT) \cite{mehta2009content} was used as a traditional FE, along with SMILY, to report the accuracy of retrieving images with the correct Gleason patterns from prostate specimens in TCGA. Our UCBMIR, as shown in Table \ref{tab:Prostate}, achieved an accuracy of $80\%$, surpassing SMILY's accuracy of $73\%$. To provide an interpretive perspective for the quantitative results, we incorporated a pre-trained VGG16 (ImageNet) as a backbone to extract histological features from the images. We added a GlobalMaxPooling2D (GMP) and two dense layers $[200, 4]$ to train the model as a classifier, in a fully-supervised manner. After training the model, we removed the last layer (Dense $(4)$) and used the remaining layers as an FE to extract $200$ features per image, which were then fed into the search engine component of UCBMIR. Figure \ref{fig:VGG.} illustrates how we integrated the pre-trained VGG16 into our CBMIR. Comparing the results shown in Table \ref{tab:Prostate} UCBMIR achieved a comparable performance as the supervised method with EV1.

We conducted experiments using SICAPv2 and Arvaniti data sets to evaluate and compare the effectiveness of our UCBMIR with respect the fully-supervised VGG16 in identifying and classifying prostate cancer images. In Table \ref{tab:Prostate_ev2}, we present the results obtained from both data sets. As shown in Tables \ref{tab:Prostate} and \ref{tab:Prostate_ev2}, our unsupervised method achieved similar results to the supervised VGG16, with a slightly lower precision score in both EV1 and EV2 (by $0.01$). This suggests that our unsupervised method is a promising approach for CBMIR in the context of prostate cancer, potentially reducing the need for manual annotation and supervision. 
\begin{table}[hpt!]
\begin{center}
\vspace{-4mm}
\caption{precision and recall at top 5, which are obtained with EV 2 on SICAPv2.}
\label{tab:Prostate_ev2}
\begin{tabular}{|c|c|c|}
\hline
 \textbf{Method} &\textbf{Data set} &\textbf{Precision}\\
\cline{1-3}                  
\hline\hline
                UCBMIR & SICAPv2 & 0.70  \\
\cline{1-3} 
                 UCBMIR & \thead{Patches normalized \\ Arvaniti (SICAP)} & 0.59  \\
\cline{1-3} 
                UCBMIR &  \thead{Patches normalized \\ Arvaniti (Panda)}& 0.68  \\

\cline{1-3} 
                  \thead{VGG16\\ \textit{(ImageNet)}} & SICAPv2 &\textbf{0.71}\\
\hline
\end{tabular}
\end{center}
\end{table}

\begin{figure*}[hpt!]
    \centering
    \subfloat[\centering \label{fig:pie chart.}]{{\includegraphics[width=.35\linewidth]{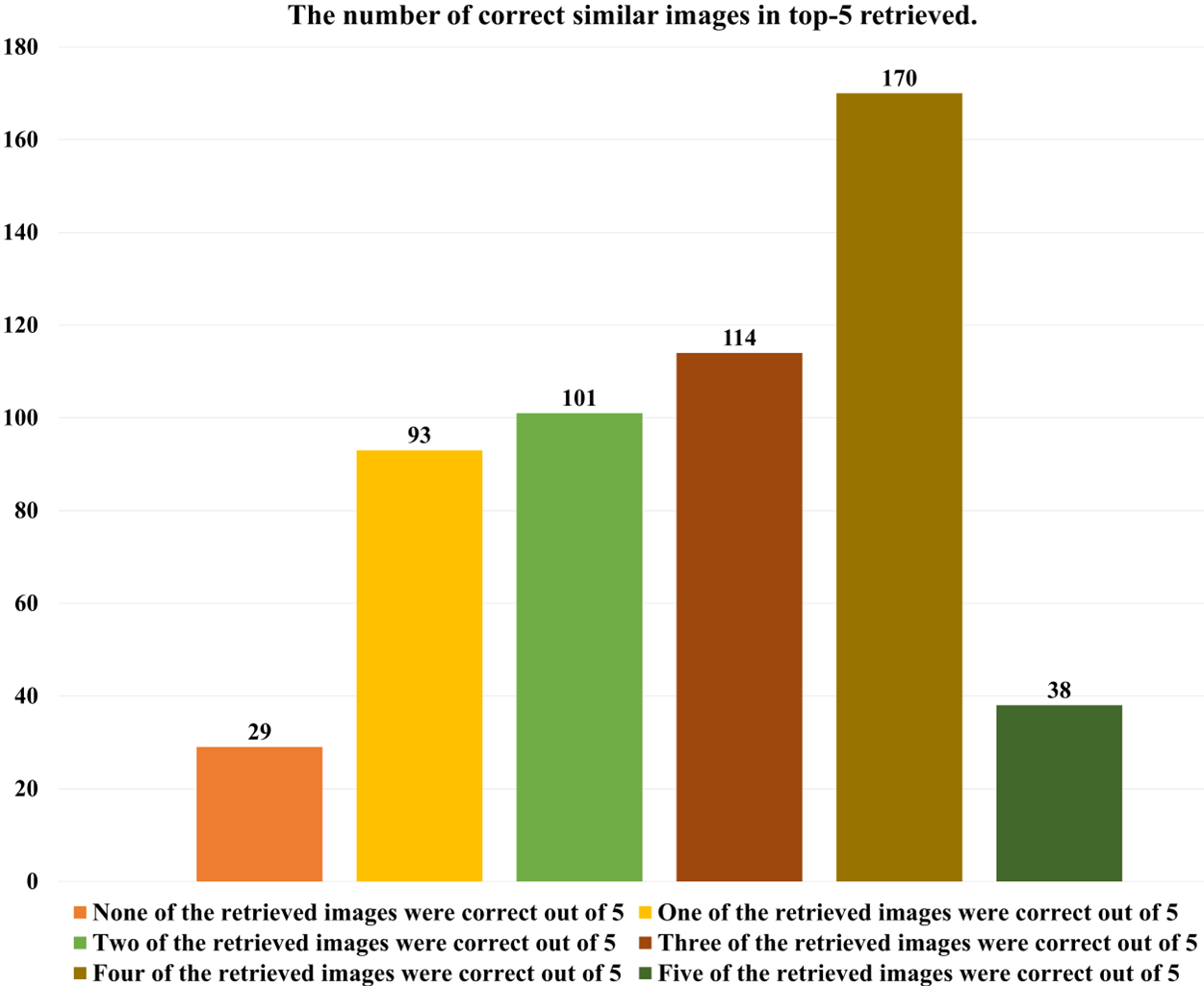} }}%
    \qquad
    \subfloat[\centering \label{fig:bar chart.}]{{\includegraphics[width=.35\linewidth]{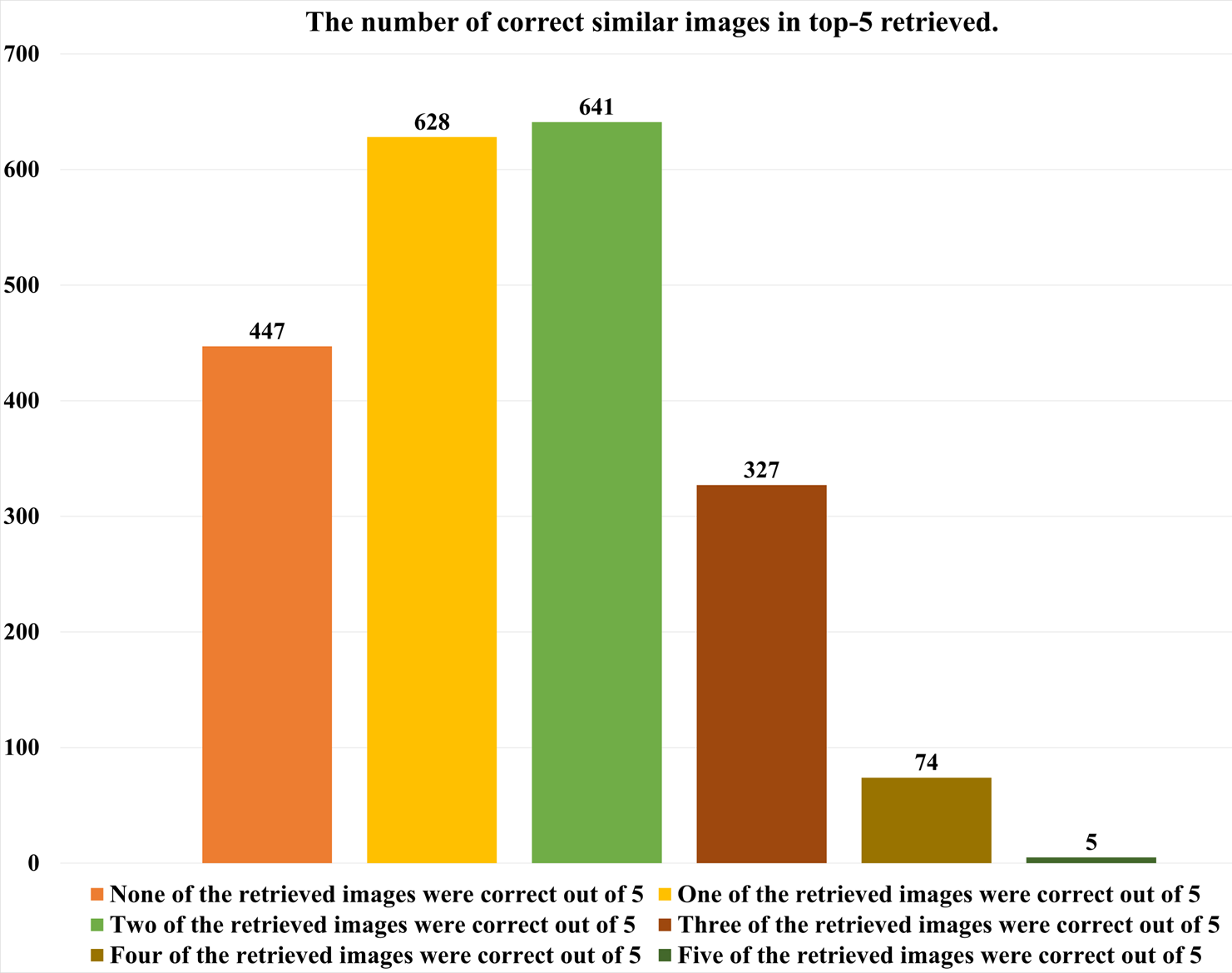} }}%
    \caption{evaluation of the UCBMIR at k = 5 on BreaKHis \ref{fig:pie chart.} and SICAPv2 \ref{fig:bar chart.}. From 2122 query images in SICAPv2, for 447 cases, the model could not find at least one correct similar images according to their labels while it retrieved two similar images at 5 top for 641 cases.}
    \label{fig:barcharts}%
\end{figure*}

In this study, Figure \ref{fig:pie chart.} and Figure \ref{fig:bar chart.} were used to presenting the results of the experiments. The bar charts were used to depict the number of similar images out of $\textit{K} = 5 $ retrieved images for BreaKHis and SICAPv2, respectively. For the BreaKHis data set, 545 images in the test set were used as query images. Based on Figure \ref{fig:pie chart.}, the model failed to find at least one similar image for 29 queries, while it could find three and four similar images for 114 and 170 queries, respectively. As for the SICAPv2 data set, the model could retrieve one similar image among the top \textit{K} for 628 image queries, according to the results shown in Figure \ref{fig:bar chart.}. 

\begin{figure*}[t!]
\begin{center}
    \includegraphics[width=.24\textwidth]{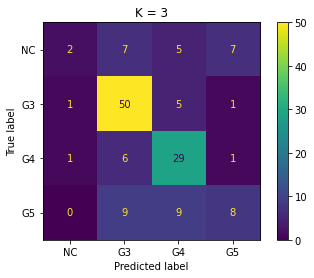}
    \includegraphics[width=.24\textwidth]{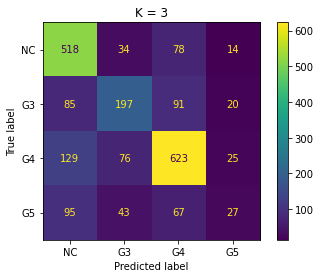}
    \includegraphics[width=.24\textwidth]{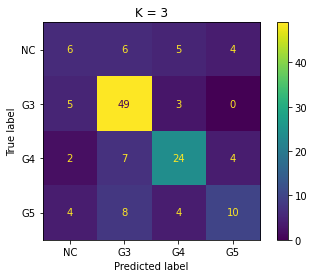}
    \\[\smallskipamount]
    \includegraphics[width=.24\textwidth]{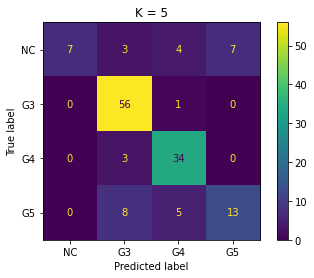}
    \includegraphics[width=.24\textwidth]{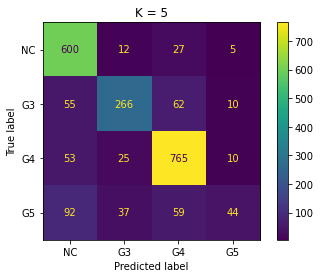}
    \includegraphics[width=.24\textwidth]{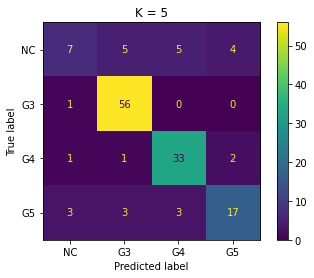}
    \\[\smallskipamount]
    \includegraphics[width=.24\textwidth]{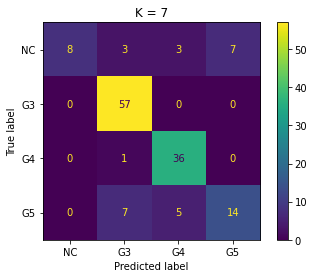}
    \includegraphics[width=.24\textwidth]{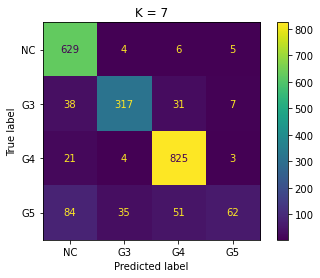}
    \includegraphics[width=.24\textwidth]{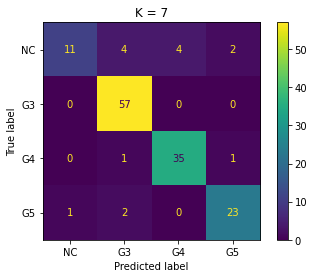}
    \caption{confusion matrix of UCBMIR on the different test cohorts at \textit{K} = 3, 5, 7. a. ARVANITI (Panda), b. SICAPv2, c. ARAVNITI (SICAP). }\label{fig:CM}
\end{center}
\end{figure*}

Due to the well-known variability between pathologists in Gleason grading and variations in histology sample preparation, it is a difficult challenge to distinguish between different grades of prostate cancer. These factors may contribute to the differences in results. Differentiating between G3 and G4 in prostate cancer requires highly experienced pathologists, takes time, and has limited inter-pathologist repeatability. However, Figure \ref{fig:CM} demonstrates the impressive ability of our UCBMIR to identify similar patterns between G3 and G4. Each row and column in Figure \ref{fig:CM} corresponds to three different values of \textit{K} in Arvaniti (Panda), SICAPv2, and Arvaniti (SICAP), respectively.

These results highlight the potential of our approach to aid in the accurate identification and CBMIR of prostate cancer images, thereby facilitating diagnosis and improving patient outcomes. Further research is needed to validate these findings on larger and more diverse data sets. Due to this, our model is also verified on an external data set with the intention of evaluating the trained model's capacity for generalization.

\begin{figure}[htp!]
\centerline{\includegraphics[width=0.68\textwidth]{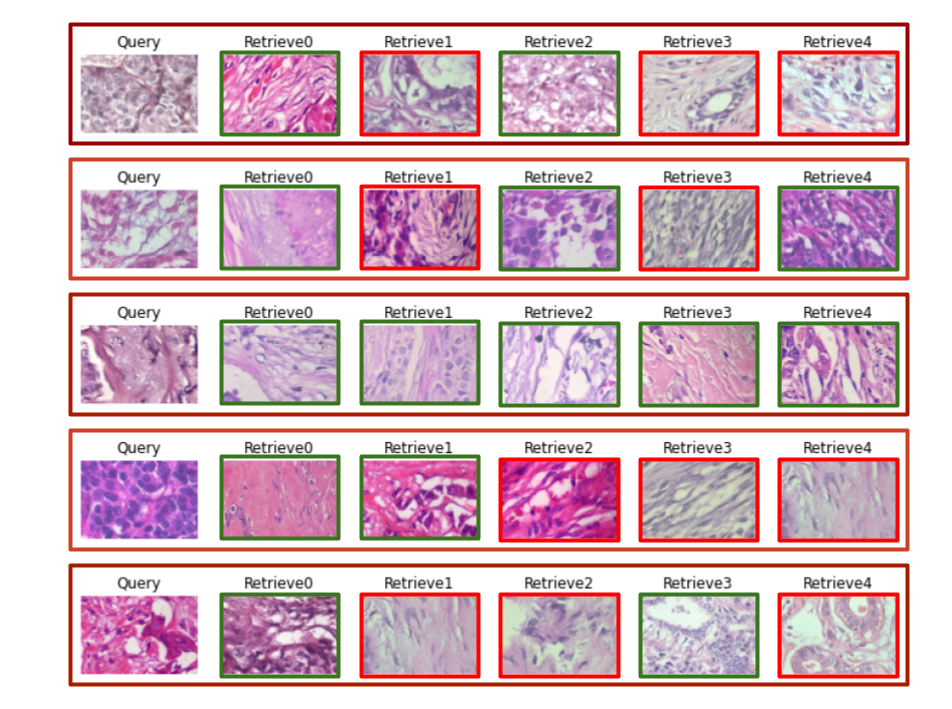}}
\caption{ the top 5 images retrieved from the BreakHis data set for five randomly selected query images, with the true and false retrieval results depicted in green and red boxes, respectively.}
\label{fig:VS_breakhis}
\end{figure}

\begin{figure}[t!]
\centerline{\includegraphics[width=0.70\textwidth]{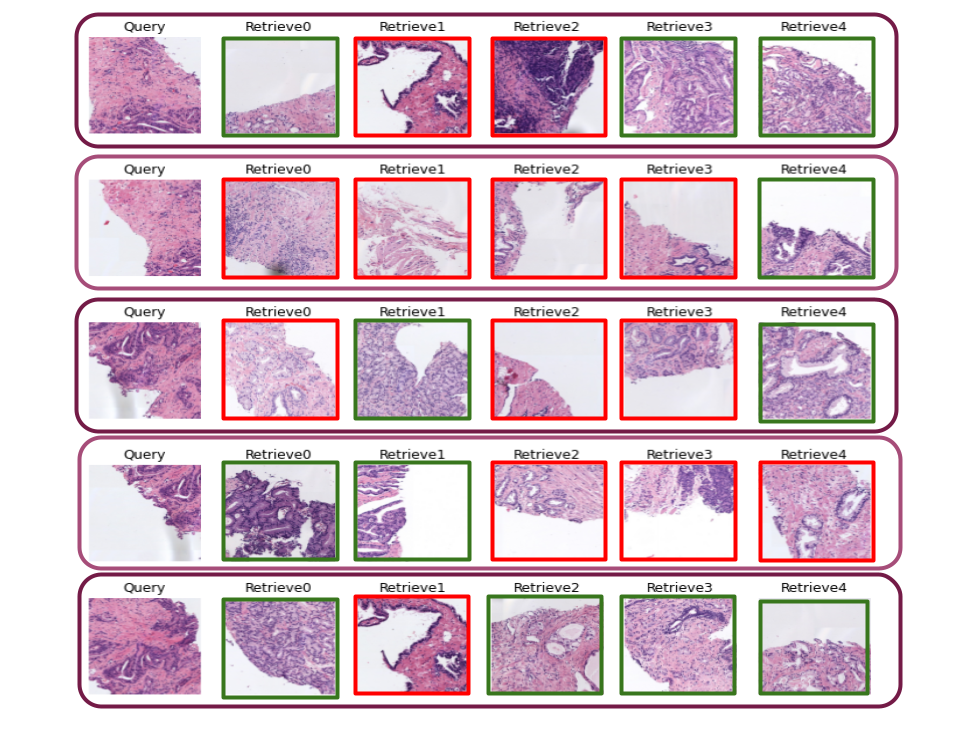}}
\caption{ the top 5 retrieved images from the SICAPv2 data set for five query images selected at random, where true and false retrieval results are respectively indicated with green and red boxes.}
\label{fig:VS_sicapv2}
\end{figure}
\begin{figure}[t!]
\centerline{\includegraphics[width=0.80\textwidth]{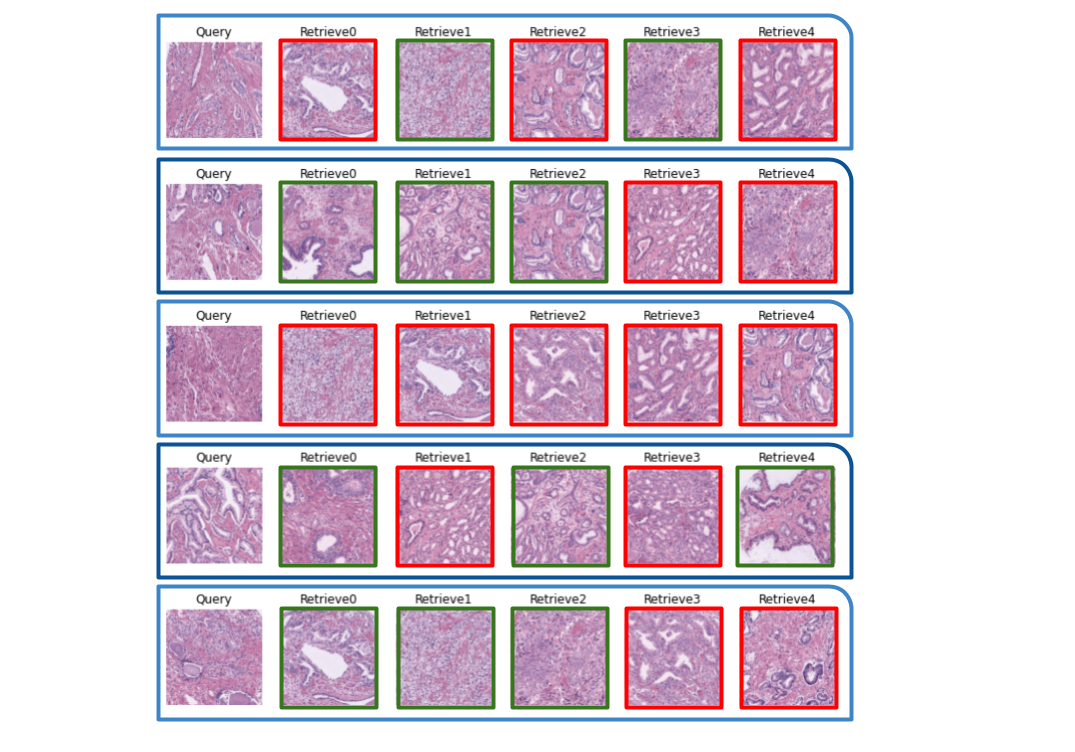}}
\caption{ the top 5 images retrieved from the Arvaniti data set, which have been normalized by SICAP, for five query images that were randomly selected. The figure visually demonstrates the results of the external validation, utilizing a well-trained model with the SICAPv2 data set and applying it in the search stage of Arvaniti. The green and red boxes respectively indicate the true and false retrieval results.}
\label{fig:VS_Arvaniti}
\end{figure}

\subsection{Validation on an external data set}
In order to validate the performance of our trained model on an external data set, we utilized the SICAPv2-trained model to make predictions on the re-sampled Arvaniti data set (normalized by SICAP). The results of this evaluation are reported in Table \ref{tab3:arvaniti}. The obtained accuracy and precision results are slightly better than those obtained from the test set on SICAPv2, while the recall is slightly lower by $0.07$. It is important to mention that this validation process is crucial in demonstrating the generalization ability of our UCBMIR beyond the original training data set. These results further confirm the robustness of our approach and its capability to provide accurate retrieval results across multiple data sets. 

\begin{table}[htp!]
\begin{center}
\caption{results of Arvaniti (normalized by SICAP) as data set for the external experiments with top \textit{K} = 5, EV 2. }
\label{tab3:arvaniti}
\begin{tabular}{|c|c|c|c|}
\hline
\textbf{\textit{K}} & \textbf{Precision} &\textbf{Recall} &\textbf{Accuracy}\\
\cline{1-4} 
\hline
5 & 0.80 & 0.73 & 0.81\\
\hline
\end{tabular}
\end{center}
\end{table}

\subsection{Visual evaluation}
The purpose of this section is to enhance comprehension of the comparison by utilizing visual aids. We have included three figures, Figure \ref{fig:VS_breakhis}, Figure \ref{fig:VS_sicapv2}, and Figure \ref{fig:VS_Arvaniti}, which showcase the results of our experiments. Each of these figures comprises rows that correspond to a random query from the test set of BreaKHis$\times400$, SICAPv2, and Arvaniti, respectively. The subsequent images in each row exhibit the top 5 retrieved images from the training set of the relevant data set. We have implemented a color-coding scheme to facilitate the interpretation of the results. In particular, a green border surrounds the correct retrieved image, which possesses the same label as the query, whereas a red border highlights the mis-retrieved images that have different labels than the query.

Figure \ref{fig:VS_sicapv2} demonstrates that one of the challenges we encountered in our experiments with SICAPv2 was the presence of a white background in the images. To determine whether the patches in SICAPv2 contained meaningful patterns for pathologists to analyze, we enlisted the help of an expert pathologist to review them.
Our pathologist confirmed that despite the presence of a white background in the images, there was still enough tissue for pathologists to evaluate and compare the patterns in the query tissue with the retrieved patches.

In addition to validating the approach of UCBMIR using an external data set and demonstrating the generalization capability of our method, we selected the Arvaniti data set for another reason: it does not have a white background. Figure \ref{fig:VS_Arvaniti} shows the top 5 retrieved images resulting from our external validation experiment, where a well-trained model with SICAPv2 was used to retrieve images for 5 random queries from the Arvaniti data set. Our external validation experiment not only validates our proposed UCBMIR for use with external data sets but also demonstrates the generalization capability of our method.

Through our visual evaluation, we aim to present a clearer understanding of the effectiveness of our approach. Observing the retrieved images alongside their labels can be useful to evaluate the performance of our method and assess its strengths and limitations. These figures are an essential component of our evaluation and will contribute significantly to understanding our methodology. Overall, this evaluation can provide valuable insights into the performance of our approach and make informed judgments regarding its effectiveness.

\subsection{Comparing UCBMIR with a classifier}

\begin{figure*}[hpt!]
    \centering
    \subfloat[\centering \label{fig:classifer}]{{\includegraphics[width=.35\linewidth]{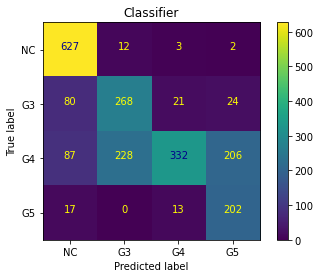} }}%
    \qquad
    \subfloat[\centering \label{fig:5top_sicap_vs}]{{\includegraphics[width=.35\linewidth]{images/5top_sicap.png} }}%
    \caption{ The confusion matrix presented in \cite{going} is displayed in Figure \ref{fig:classifer}, while the matrix for the retrieved labels is shown in Figure \ref{fig:5top_sicap_vs}. It can be observed that the UCBMIR model results in less conflict between the challenging grades (G3 and G4) compared to the classifier.}
    \label{fig:classification_CBMIR_cm}%
\end{figure*}

CBMIR and classification are two different approaches in medical image analysis. CBMIR aims to retrieve similar images from a database based on the content features of a query image, while classification aims to categorize images into pre-defined classes or labels. The only mutual output of the classification tool and CBMIR is the output labels corresponding to the output patches. 
In the above sections, it is mentioned that UCBMIR achieved comparable results with supervised CBMIR techniques, proven by the reported accuracies in Table \ref{tab:Prostate} and Table \ref{tab:Prostate_ev2}. Regarding comparing the predicted labels in terms of retrieving similar images belonging to the same cancer type, we provide Table \ref{tab:CBMIR_VS_Classification}. This table compares the accuracy of UCBMIR with a classifier in \cite{going} in both the validation and the test set of SICAPv2.

the proposed unsupervised CBMIR model was found to be highly effective in distinguishing between different cancer grades, especially between the challenging Gleason grades G3 and G4. This observation was evident from the confusion matrix shown in Figure \ref{fig:classification_CBMIR_cm}. The proposed method's success can be attributed to its ability to identify and utilize subtle features and patterns in the images that may be missed by human observers or conventional supervised models. 

\begin{table*}[htp!]
\centering
\caption{shows a comparison between the performance of UCBMIR with EV 1 and the classification introduced in \cite{going}. SICAPv2 is the the data set under study.}
\label{tab:CBMIR_VS_Classification}
\begin{tabular}{|c|c|c|}
\hline
                  Data set & Model & Accuracy \\ \hline
\multirow{2}{*}{Validation set} & UCBMIR & \textbf{0.83}\\ \cline{2-3} 
                  &Classification\cite{going}  & 0.76 \\ \hline
\multirow{2}{*}{Test set} & UCBMIR & \textbf{0.79}\\ \cline{2-3} 
                  &Classification\cite{going} & 0.67 \\ \hline

\end{tabular}
\end{table*}

\section{Conclusion}
In summary, this paper introduces a highly qualified Unsupervised CBMIR (UCBMIR) model that can be used for both binary and multi-class data sets. The model was evaluated on three different data sets, as well as an external validation set. Using two most-used evaluation techniques, the proposed method achieved 79\% and 70\% precision in EV 1 and EV 2, respectively on SICAPv2 as a multi-class data set. Notably, the unsupervised method was able to differentiate between challenging Gleason grades of prostate cancer. In addition to numerical evaluation, visual assessments were conducted to demonstrate the effectiveness of the UCBMIR. The results show that UCBMIR has good generalizability and can be effectively applied to other types of cancer. UCBMIR has the potential to improve laboratory productivity, increase pathologists' diagnostic confidence, and contribute to the advancement of cancer diagnosis and treatment. 

The UCBMIR model not only addresses the needs and challenges of pathologists but also could address the problem of engineers who face a lack of sufficient images for training models. Future research in this field could build on these findings and further enhance the performance of CBMIR models for cancer diagnosis.
\section{Future work}
In the world of CBMIR, there is a vast range of possibilities for enhancing and optimizing laboratory productivity. With a large archive of diagnosed patients and corresponding data, including images and treatment and monitoring reports, it should be possible to identify and retrieve images that are either anatomically or pathologically similar to the biopsy sample of the patient being examined, as well as the annotated data for each case. CBMIR has the potential to be applicable to many types of cancer, which would increase its utility.

Furthermore, pathologists' reports contain the medical knowledge of many other pathologists for similar cases, making them a treasure trove of high-quality diagnostic information. In the future of CBMIR, it may be possible to make the raw information directly available to the pathologist or to merge the important information in retrieved reports. This would make the diagnosis process more efficient, accurate, and informative for both the pathologist and the patient. Additionally, expanding the use of CBMIR to other types of medical imaging and diagnostic data could provide valuable insights for a range of medical specialties.

\section{Acknowledgement}
I sincerely thank Umay Kiraz and Andrés David Mosquera Zamudio for their collaboration and efforts in image annotation. Their expertise and dedication were invaluable, and I am grateful for the opportunity to work with them.

This research is financially supported by the Marie Skłodowska-Curie grant agreement No. 860627 from the European Union's Horizon 2020 research and innovation program, as part of the CLARIFY Project. Additionally, the contributions of Adrián Colomer have received support from the ValgrAI – Valencian Graduate School and Research Network for Artificial Intelligence, as well as the PAID-PD-22 program of the Generalitat Valenciana and Universitat Politècnica de València.
\bibliography{sample}
\section*{Author contributions statement}
Zahra Tabatabaei was responsible for the conception of the experiments, analysis of results, validation, writing of the original draft, and study design. At the same time, all other authors reviewed the manuscript.

\end{document}